\documentclass[12pt]{article}
\usepackage{graphicx}

\begin{document}
 \title{A time representation}
\author{Lucas Lamata and Juan Le\'on\footnote{lamata@imaff.cfmac.csic.es, leon@imaff.cfmac.csic.es} \\
 {\small Instituto de Matem\'aticas y F\'{\i}sica Fundamental,
 CSIC,} \\\small{Serrano 113-bis, 28006 Madrid, Spain}}
 \maketitle
\begin{abstract}
The paper contains a proposal for an energy and time
representation. We construct modes that correspond to fuzzy
distributions around discrete values of energy or time. The modes
form an orthogonal and complete set in the space of square
integrable functions. Energy and time are self adjoint in the
space spanned by the modes. The widths of the modes are analyzed
as well as their energy-time uncertainty relations. The lower
uncertainty attainable for the modes is shown. We also show times
of arrival for massless particles.
\end{abstract}
 \section*{The Pauli theorem revisited}

Two arbitrary states of an elementary system can be transformed into
each other by symmetry operations. This opens the door to express
what can be observed of the system, i.e. the system properties, in
terms of the generators of these symmetries. In the case of the
Poincare group they are the momenta $\hat{P}^{\mu}$ and the angular
momenta  and boosts $\hat{M}^{\mu \nu}$. For the system to be
elementary they are constrained by the mass shell condition
$\hat{P}^2 = m^2$ and by the spin condition $\hat{W}^2 = m^2
s(s+1)$. (The Pauli-Lubanski vector is defined as $\hat{W}_{\mu} =
\epsilon_{\mu \nu \alpha \beta} \hat{P}^{\nu} \hat{M}^{\alpha
\beta}$, with $\hbar =1$ in this paper unless otherwise specified).

Notice now the conjunction of both, the four vector character of
the momenta on one side, with the necessity of introducing
conjugate operators to the three-momentum to formulate dynamics on
the other. This calls for the introduction of a four vector
operator $\hat{Q}^{\mu}$ conjugate to $\hat{P}^{\mu}$ such that
\begin{equation}
[\hat{P}^{\mu},\hat{Q}^{\nu}] = i  g^{\mu \nu} \label{a}
\end{equation}

 Regretfully, this is not possible to attain. The reason is that
 $\hat{Q}^{\mu}$ would produce translations in the momentum. If
 $\hat{P}^{\mu}$ is defined on the mass shell, then
\begin{equation}
\hat{P'}^{\mu} = \exp{(-i \delta p \hat{Q})} \, \hat{P}^{\mu} \,
\exp{(i \delta p \hat{Q})}\, =\, \hat{P}^{\mu} + \delta p^{\mu}
\label{b}
\end{equation}
and, no matter the value of the  four vector parameter $\delta p$,
$\hat{P'}^2 \neq m^2$. The transformations generated by $\hat{Q}$
pull the momentum $\hat{P}$ out of the particle mass shell.

The problem is independent of the form of dynamics in use. It has
far reaching consequences. First, it is necessary to abandon a
four vector $\hat{Q}^{\mu}$ and hence the hope of building  a
covariant form dynamics in terms of momenta and positions~\cite{teller}. Then,
with time and position demoted to the role of mere parameters, it
is necessary to introduce quantum fields with creation and
annihilation operators playing the role of conjugate pairs. To our
accustomed eyes they look like the appropriate recipe for particle
number non conservation~\cite{crew} but, Is it so? At least two important
lessons remain: 1. That time is along the direction that would be
conjugate to the solution of the mass shell constraint and 2. That
time runs due to the constraint.

With the usual non covariant choice, where the mass shell
condition reads as $\hat{P}^0 = H(\hat{\bf{P}})$, the problem
turns into a version of the Pauli theorem~\cite{pauli}, namely
\begin{equation}
\hat{H'} = \exp{(-i \delta E \hat{T})} \, \hat{H} \, \exp{(i
\delta E \hat{T})}\, =\, \hat{H} + \delta E,\; \mbox{with}\,
\delta E \in \mathbf{R} \label{c}
\end{equation}
This implies that the spectrum of $\hat{H}$ has to be the real
line, something that runs again the existence of the physically
necessary ground state. So, in the time form of dynamics the
problem looks like the incompatibility of $\hat{T}$ with the
boundedness of the Hamiltonian. On the other hand, if $\hat{T}$ is
not selfadjoint, then $[\hat{H},\hat{T}] \neq i$ and there are way
outs from the problem.

Perhaps the simplest case for an observable time is the time of
arrival (TOA) of a particle at a certain position in one space
dimension~\cite{muga}. Classically $t(x)=m (x-q)/p$ where $q,p$ are the
dynamical variables of phase space (initial position and momentum)
and $x$ the arrival position. Obtaining the quantum version
$\hat{t}(x)$ of this TOA is complicated due to the presence of
$1/\hat{p}$ and to operator ordering. It is well known that there
is no selfadjoint $\hat{t}(x)$, its most symmetrical expression
being
\begin{equation}
\hat{t}(x) = -\exp(-ix\hat{p})\,\sqrt{\frac{m}{\hat{p}}}\,\,
\hat{q}\,\sqrt{\frac{m}{\hat{p}}}\,\exp(ix\hat{p}) \label{d}
\end{equation}
The eigenvectors of this operator, $|t,x,s\,\rangle$ can be given
in  the  momentum representation where $1/\hat{p}$ is easier to
deal with, as
\begin{equation}
\langle p\,|t,x,s\,\rangle\, =\, \theta(sp)\,\sqrt{\frac{|p|}{m}}
\,\exp(i \frac{p^2}{2m} \,t)\, \langle p\,|x\,\rangle  \label{e}
\end{equation}
The degeneration parameter $s$ can take the  values +1 for right
movers, -1 for left movers.This is the only residue left at one
space dimension of the continuous manifold of directions present
for higher D.

The bad news come in the form of nonorthogonality of the time
eigenstates. This is a consequence of the fact that the
Hamiltonian is bounded from below: $\sigma(\hat{H})=\,
[E_0,\infty)$ with $E_0=0$. By using  (\ref{e}) we get

\begin{eqnarray}
\langle t,x,s\,|t',x,s'\,\rangle\, &=\,& \frac{1}{2\pi}\,
\delta_{ss'}\,\int_{E_0=0}^{\infty}  \, \exp(i E (t-t')) =
\nonumber\\ \frac{1}{2\pi}\,\delta_{ss'}\,\lim_{\epsilon
\rightarrow 0^+} \frac{i}{t'-t+i\epsilon} &= &
\frac{1}{2}\,\delta_{ss'}\,(\delta(t-t')-\frac{i}{\pi}
{\mathcal{P}} \frac{1}{t-t'})  \label{f}
\end{eqnarray}

To get orthogonality it is necessary to move $E_0$  to $-\infty$
something that resembles the Weisskopf-Wigner trick for resonances.
In any case it is clear that the problem source is in
$\sigma(\hat{H})\neq \sigma(\hat{t})$.

Marolf devised a procedure, used in ref~\cite{tate}, to surmount
nonorthogonality. The idea is to replace $\hat{t}$ and $\hat{H}$
by ``regularized" expressions that avoid the difficulties that
arise when $p \rightarrow 0$, namely, to introduce
\begin{equation}
f_{\epsilon} (p) = \{ \frac{m}{|p|} \,\mbox{if}\, |p| >
\epsilon,\,\, \mbox{else}\, \frac{m |p|}{\epsilon^2} \}\;
\mbox{with}\,\epsilon \,\mbox{small and positive} \label{g}
\end{equation}
Then,
\begin{equation}
\hat{t}(x) = -\exp(-ix\hat{p})\,\sqrt{f_{\epsilon}(\hat{p})}\,\,
\hat{q}\,\sqrt{f_{\epsilon}(\hat{p})}\,\exp(ix\hat{p}) \label{h}
\end{equation}
an expression tailored for the momentum representation where
\begin{equation}
\langle p\,|t,x,s\,\rangle_{\epsilon}\, =\,
\theta(sp)\,\frac{1}{\sqrt{f_{\epsilon}(p)}} \,\exp(i
E_{\epsilon}(p) \,t)\, \langle p\,|x\,\rangle  \label{i}
\end{equation}
The ``regularized energy" is given by
\begin{equation}
E_{\epsilon}(p)\,=\,\int_{\pm \epsilon}^p
\frac{dp'}{f_{\epsilon}(p')}\,=\, \{E(p)-E(\epsilon)\,\,
\mbox{if}\, |p| > \epsilon,\,\,
\mbox{else}\,E(\epsilon)\,\ln(\frac{E(p)}{E(\epsilon)})\}
  \label{j}
\end{equation}
Thus, while $E(p)=0$ when $p=0$, $E_{\epsilon}(p)\rightarrow
-\infty$ as $p \rightarrow 0$. This solves the problem by removing
the lower bound in $\sigma(\hat{H})$. It is straightforward to show
that
\begin{equation}
_{\epsilon}\langle t,x,s\,|t',x,s'\,\rangle_{\epsilon}\, =\,
\delta_{ss'}\delta(t-t') \label{k}
\end{equation}

The procedure works fine for the case of free particles. With
minor obvious modifications, it works equally well for
relativistic particles and for any number of space dimensions.
However, it is only suitable for those cases where momentum
remains constant. So, its very definition brings this procedure to
a dead end.

Soon after the publication of ref~\cite{tate}
Giannitrapani~\cite{gianni} observed that  the time of arrival was
an instance of generalized observable endowed with a probabilistic
interpretation as a positive operator valued (POV) measure. In
fact, the time of arrival eigenstates form a complete set:
\begin{equation}
\langle p\,|\, \{\sum_s \int  dt\, |t,x,s\,\rangle\,\langle
t,x,s\,|\,\}\,|p'\,\rangle \,=\,\langle p\,|p'\,\rangle \label{l}
\end{equation}
Not being orthogonal they can not constitute a projector valued
measure, but -- as pointed out in ref~\cite{gianni} -- nothing
prevents from using them to construct a POV measure. Giannitrapani
showed that the probability that the time of arrival at $x$ of the
state $\rho$ be in the range $[T,T']$ is
\begin{equation}
P_x\left([T,T']\right)\,=\,Tr\left[\rho\,\left(\sum_s
\int_{T}^{T'}\, dt\, \,|t,x,s\,\rangle\,\langle
t,x,s\,|\right)\right]\label{m}
\end{equation}
The mean value and variance of the TOA are given in~\cite{gianni},
where the problems arising from nonorthogonality when trying to
obtain kinematical uncertainty relations between energy and time
are discussed. The paper also analyzes the free particle TOA
operator of (\ref{d}) acting on the domain of infinitely
differentiable functions  over the compact subsets of values of $p
\,\in\,\mathbf{R}-\{0\}$. In this case the variance of $\hat{t}$
turns out to be computable as for ordinary observables and the
Heisenberg uncertainty relation holds. To summarize: Endowed with
the POV measure interpretation, the TOA became an useful
instrument appropriate to extract physical information from one of
the primary laboratory events: ``when" a detector clicks. No
wonder this is the customary approach in the current literature.

\section*{From eigenvectors to wave packets}
Wigner introduced~\cite{wigner} in 1932 a distribution at midway
between  the position and the momentum representation with the aim
of describing particle properties in phase space. The non
positivity of the distribution was an obstruction to its use as a
probability distribution. Later on, Husimi~\cite{husi} introduced sets of
minimal uncertainty states $|q\,p\rangle$ in position and
momentum:
\begin{equation}
\langle x|q\,p\rangle\,=\,(2 \pi \sigma^2)^{-1/4}
\exp(-\frac{(x-q)^2}{4\sigma^2})\, \exp(i px) \label{n}
\end{equation}
The distribution is centered at the point $(q,p)$. Given a system
in an arbitrary state $|\psi\,\rangle$, the probability that the
system occupy a region in phase space centered at $(q,p)$ of half
widths $(\sigma,1/2\sigma)$ is given by
$\rho_H(q,p)\,=\,(2\pi)^{-1}|\langle q\,p\,|\psi\,\rangle|^2$

We now return to the time of arrival of a free particle at a
specific position in one space dimension. We learnt the
difficulties that arise in this seemingly simple problem. They can
be articulated through the Pauli theorem in two complementary ways:
The impossibility of finding a conjugate pair of time and energy
selfadjoint operators or, the difficulty to connect the two
different spectra $\sigma(\hat{t})=\mathbf{R}$ and
$\sigma(\hat{H})=\mathbf{R_+}$. To avoid them, we will follow a
procedure that resembles Husimi's.

We start by introducing two functions $g_{\nu}(t)$ and
$f_{\nu}(\omega)$ defined over the real line $t
\in\mathbf{R}\,=\,(-\infty,+\infty)$ and the positive real line
$\omega\in\mathbf{R}_+\,=\,(0,+\infty)$ respectively. We take
$\nu$ as a positive integer $\nu\,>\, 1$. Also, we assume
$g_{\nu}$ and $f_{\nu}$ to be Fourier transforms of each other.
Finally,
\begin{equation}
g_{\nu}(t)\,=\,
\frac{1}{(\beta+i\,t)^{\nu}},\;\;f_{\nu}(\omega)\,=\,
\frac{\sqrt{2\pi}}{\Gamma(\nu)}\,\omega^{\nu-1}\,\exp(-\beta
\omega) \label{o}
\end{equation}

These functions on which t (respectively $\omega$) act
multiplicatively, present the following nice property under
Fourier transformation:
\begin{eqnarray}
t\,g_{\nu}(t)&\leftrightarrow&-i\,\frac{df_{\nu}(\omega)}{d\omega}\nonumber\\
i\,\frac{dg_{\nu}(t)}{dt}&\leftrightarrow&\omega\,f_{\nu}(\omega)
\label{p}
\end{eqnarray}
We point out that for $\nu>1$ both derivatives act on these
functions as selfadjoint operators ($f_{\nu}(0)=0$). It is also
remarkable that $g_{\nu}\in L^2(\mathbf{R})$ and $f_{\nu}\in
L^2(\mathbf{R_+})$. So, they can be given a probabilistic
interpretation. Finally, to the canonical pair $(t,i\,\frac{d}{dt})$
acting on $t$ (the would be ``Time Representation") corresponds the
unitarily equivalent pair $(-i\,\frac{d}{d\omega},\, \omega)$ acting
on  $\omega$ (the would be ``Energy Representation").

This looks like a promising starting point for constructing true
conjugate representations for time and energy. The obstruction is
the need of completeness of the $g_{\nu}$ in $\mathbf{R}$ and of
the $f_{\nu}$ in $\mathbf{R}_+$ to build systems of generators for
square integrable functions. This is a necessary condition to
arrive at a POV measure. If in addition we find orthogonality,
then we would reach a PV measure. Notice however, that these
measures would correspond to fuzzy distributions around a central
value, in much the same way that the Husimi Gaussian wave packets
do. This can be seen with our $g_{\nu}$ and $f_{\nu}$, which are
not eigenstates of time or energy.
\begin{equation}
t\,g_{\nu}(t)\,=\,i\beta \,g_{\nu}(t)-\,i\,g_{\nu-1}(t),\;\;
\omega\,f_{\nu}(\omega) \,=\,\nu\,f_{\nu+1}(\omega) \label{q}
\end{equation}
The construction of eigenfunctions is hopeless at this stage. For
instance, be $\psi(\omega)\,=\, \sum_{\nu \geq \nu_0} c_{\nu}
f_{\nu}(\omega)$ a putative eigenfunction, so that $\omega
\,\psi(\omega)\,=\, \lambda\,\psi(\omega)$. Then
$\psi(\omega)\propto \delta(\omega-\lambda)$;  expanding this in
the $f_{\nu}(\omega)$ would at least require that they form a
complete set, which is not the case. The power of this simple
example is that it signals the way to proceed.

\section*{Energy and time representations}
Orthogonal polynomials are a useful tool to solve a variety of
problems in physics. The Laguerre polynomials $L_n^{\alpha}(x)$
constitute a set of orthogonal polynomials on the interval
$(0,\infty)$ with weight $x^{\alpha}\, e^{-x}$. Accordingly, we
can define the set of orthogonal functions
\begin{equation}
\varphi_n^{\alpha}(\omega),=\,\{0\; \mbox{if}\; \omega<\,0,\;
{\mbox{else}}\,\,
 c_n^{\alpha}\,
(\frac{\omega}{\omega_0})^{\alpha/2}\,e^{-\omega/2\omega_0}\,L_n^{\alpha}
(\omega/\omega_0)\}
\label{r}
\end{equation}
where
\begin{equation}
 c_n^{\alpha}\,=\,\left(\frac{\Gamma(n+1)}{\omega_0
\Gamma(\alpha+n+1)}\right)^{1/2}\,\,L_n^{\alpha}(x)\,=\,\sum_{m=0}{n}
(-1)^m \left({n+\alpha \atop n-m
}\right)\,\frac{x^m}{m!}\label{ra} \end{equation}
 and $\omega_0$ is a short of
width of the exponential distribution that also serves to keep
dimensions right (still, $\hbar=1$). This definition, provides a
discrete denumerable set of modes that constitute a basis for all
the functions belonging to $L^2(0,\infty)$. These functions
comprise all functions with positive frequencies (i.e. $\omega>0$)
that can also be associated to probabilities (in fact,
$\int_0^{\infty} d\omega |\varphi_n^{\alpha}(\omega)|^2\,=\, 1$).
The orthogonality and completeness relations read explicitly as
\begin{equation}
\int_0^{\infty} d\omega\,
\overline{\varphi_n^{\alpha}(\omega)}\,\,\varphi_{n'}^{\alpha}
(\omega)\,=\,\delta_{nn'}\;\mbox{and}\;
\sum_n\,\,
\,\varphi_{n}^{\alpha}(\omega)\,\,\overline{\varphi_n^{\alpha}
(\omega')}\,=\delta(\omega-\omega')\label{s}
\end{equation}
\begin{figure}[h]
\begin{center}
\includegraphics{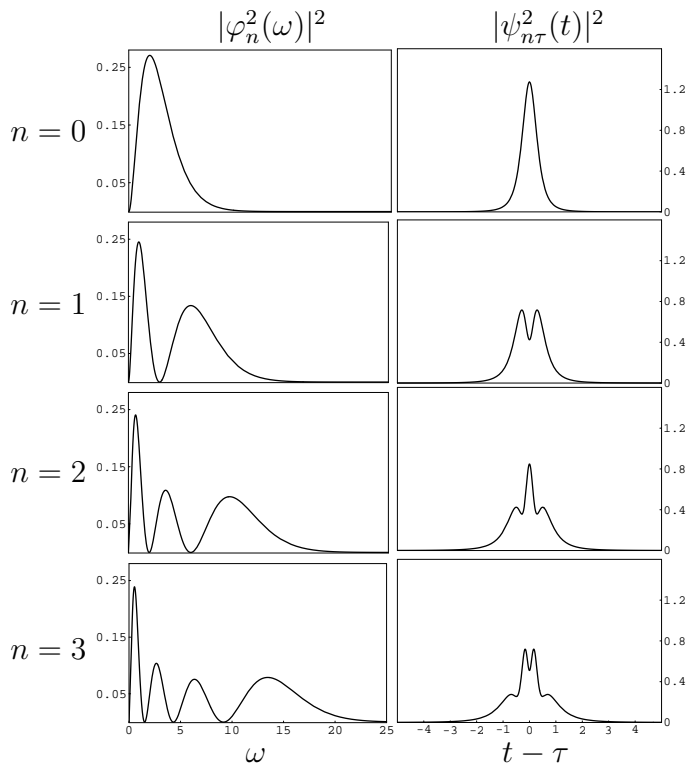}
\caption{The lowest modes for $\alpha=2$. $\omega$ is given in units
of $\omega_0$ and  $t-\tau$ in units of $\omega_0^{-1}$. The number
of maxima for each mode is $n+1$. The mode widths grow with
increasing $n$.}
\end{center}
\end{figure}
The physical information contained in the modes is readily
obtained. In this representation, the Hamiltonian, in spite of its
simple form $\hat{H}\,=\,\omega$ is  a non-diagonal matrix with
elements given through
\begin{equation}
\hat{H}\,|\,n\,\rangle\,=\,\sum_{m=n-1}^{n+1}\,d^1_{mn} \,\omega_0
\,|\,m\,\rangle \label{t}
\end{equation}
In the same way
\begin{equation}
\hat{H}^2\,|\,n\,\rangle\,=\,\sum_{m=n-2}^{n+2}\,d^2_{mn}
\,\omega_0^2 \,|\,m\,\rangle \label{u}
\end{equation}
The coefficients $d_{mn}$'s are computable by standard methods. Notice that
$d^2_{mn}$ is not $(d^1_{mn})^2$. The average value of the energy
in the n-th mode is
\begin{equation}
\langle\,n\,|\,\hat{H}\,|\,n\,\rangle\,=\,d^1_{nn}\,=\,
(\alpha+2n+1)\, \omega_0 \label{v}
\end{equation}
and the average of the squared Hamiltonian
\begin{equation}
\langle\,n\,|\,\hat{H}^2\,|\,n\,\rangle\,=\,d^2_{nn}\,=\,
[(n+1)(\alpha+n+1)+(\alpha+2n+1)^2+n(\alpha+n)]\,
 \omega_0^2 \label{w}
\end{equation}
Finally, the variance of $\hat{H}$, $(\Delta H)_n^2$ in the  n-th
mode is given by
\begin{eqnarray}
(\Delta H)_n^2\,&=&\,\langle\,n\,|\,\hat{H}^2\,|\,n\,\rangle\,-
\,\langle\,n\,|\,\hat{H}\,|\,n\,\rangle^2\,=\,d^2_{nn}-(d^1_{nn})^2
\nonumber\\\,&=&\,[(n+1)(\alpha+n+1)+n(\alpha+n)]\omega_0^2\,
\label{x}
\end{eqnarray}
Another question to investigate are the eigenvalues $E$ and
eigenfunctions $\Psi_E$ of the Hamiltonian. $\Psi_E$ can be
expanded in terms of the modes as $\Psi_E\,=\,\sum_n c_n
\,|\,n\,\rangle$ which translates into the set of linear equations
$\sum_n\,\omega_0\,d^1_{mn}\,c_n\,=\,E\,c_m$  left to the interested reader as an exercise.

We take the next step forward and define wave packets centered at
the point $(\tau,E)$ in time energy space. To simplify the
discussion we assume that $E$ is the average energy in the n-th
mode $E=(\alpha+2n+1)\omega_0$. We assume -- for our purposes here --  that we can trade $(\tau,n)$
by $(\tau,E)$. Finally, the wave packet we are looking for is:
\begin{equation}
\varphi_{n \tau}^{\alpha}(\omega)\,=\,e^{i\omega
\tau}\,\varphi_{n}^{\alpha}(\omega)\label{y}
\end{equation}

The meaning of this time $\tau$ just introduced can best be explored
in the time representation. The Fourier transform of the modes
$|n,\tau\rangle$ bring them from the energy to the time
representation that is
\begin{equation}
\psi_{n
\tau}^{\alpha}(t)\,=\,\frac{1}{\sqrt{2\pi}}\,\int_{0}^{\infty}
d\omega \,e^{-i\omega
(t-\tau)}\,\varphi_{n}^{\alpha}(\omega)\label{z}
\end{equation}
\begin{figure}[h]
\begin{center}
\includegraphics{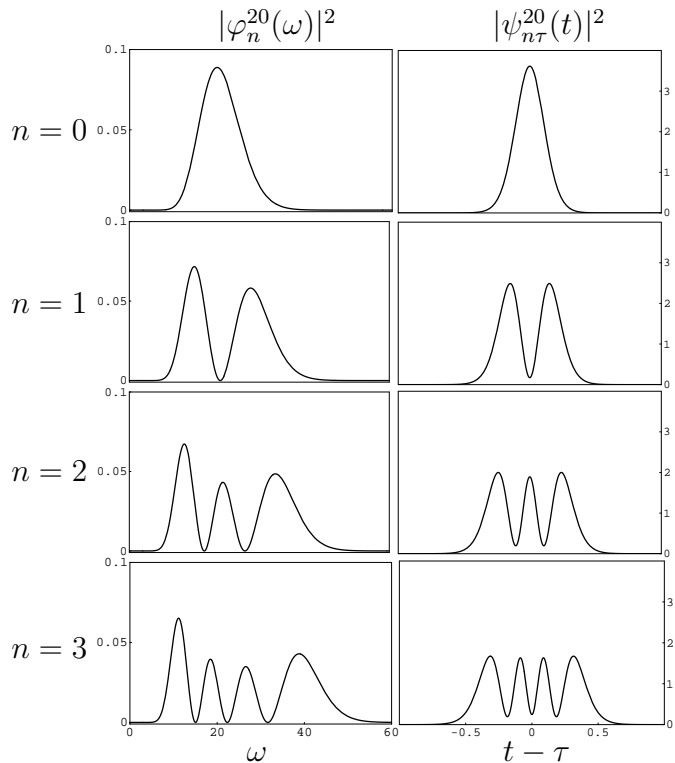}
\caption{The effect of  increasing $\alpha$. The modes plotted
here are for $\alpha=20$. Notice that here the width in $t$ is  one order of magnitude smaller than in Fig. 1}
\end{center}
\end{figure}

Notice that $\psi_{n \tau}^{\alpha}(t)\,=\,\psi_{n
0}^{\alpha}(t-\tau)$ by construction. Also, as the $\varphi_{n
\tau}^{\alpha}(\omega)$ are a basis in $\mathbf{R}_+$ they
constitute a basis for $L^2( \mathbf{R})$. So, we conclude that
they form a system of imprimitivity leading to a PV measure. This
is valid for both, the $\omega$ and the $t$ representations as
they are unitarily equivalent.

The explicit expression of the modes in the $t$ representation is
quite cumbersome, but throws some light on their physical content
and on the way that the two different spectra are connected. From
ref~\cite{tables} we learn that the integral in (\ref{z}) can be
carried out giving:
\begin{eqnarray}
\psi_{n \tau}^{\alpha}(t)\,&=&\,\frac{\omega_0
}{\sqrt{2\pi}}\,c_n^{\alpha}\,\frac{\Gamma(\alpha/2+1)\Gamma(\alpha+n+1)
}{n! \Gamma(\alpha +1)}\nonumber\\ &\times&\frac{_2F_1\left(
-n,\alpha/2+1;\alpha+1;(1/2+i\omega_0
(t-\tau))^{-1}\right)}{(1/2+i\omega_0 (t-\tau))^{\alpha /2 +1}}
\label{aa}
\end{eqnarray}
Being the first entry a negative integer $-n$, the hypergeometric
function terminates. Explicitly
\begin{eqnarray}
_2F_1\left( -n,\alpha/2+1;\alpha+1;(1/2+i\omega_0
(t-\tau))^{-1}\right)\,&=&\nonumber\\
\sum_{m=0}^{n}\,\frac{(-n)_m (\alpha/2+1)_m}{(\alpha+1)_m \,\,
m!}\,(1/2+i\omega_0 (t-\tau))^{-m}& &\label{ab}
\end{eqnarray}
where each term in the sum in (\ref{ab}) comes from the
corresponding term in the sum in (\ref{ra}).

Due to the fact that $\overline{\psi_{n 0}^{\alpha}(t)}=\psi_{n
0}^{\alpha}(-t)$, we have $\langle
n,\tau|\hat{T}|n,\tau\rangle=\tau$. To compute $\langle
n,\tau|\hat{T}^2|n,\tau\rangle$, we consider the derivative of the
mode $\varphi_n^{\alpha}(\omega)$($\omega_0=1$):
\begin{eqnarray}
\frac{d}{d\omega}\varphi_n^{\alpha}(\omega)=\left(\frac{\alpha}
{2\omega}-\frac{1}{2}\right)\varphi_n^{\alpha}(\omega)-
\frac{c_n^{\alpha}}{c_{n-1}^{\alpha}}\varphi_{n-1}^{\alpha}
(\omega)\Theta(n,0),\label{ac}
\end{eqnarray}
where we define $\Theta(n,0)=1$ when $n>0$ and zero otherwise. On
the other hand, the Laguerre polynomials verify
\begin{equation}
L_n^{\alpha}=\sum_{m=0}^n L_m^{\alpha-1},\label{ad}
\end{equation}
and thus
\begin{eqnarray}
\frac{d}{d\omega}\varphi_n^{\alpha}(\omega)=\sum_{l=0}^n
N_{nl}^{\alpha}\varphi_l^{\alpha-2}(\omega)-\frac{1}{2}
\varphi_n^{\alpha}(\omega)
-\frac{c_n^{\alpha}}{c_{n-1}^{\alpha}}\varphi_{n-1}^{\alpha}
(\omega)\Theta(n,0),\label{ae}
\end{eqnarray}
where
\begin{equation}
N_{nl}^{\alpha}=\frac{\alpha}{2}\frac{c_n^{\alpha}}{c_l^{\alpha-2}}(n-l+1).\label{af}
\end{equation}
The matrix element $\langle n,\tau|\hat{T}^2|n,\tau\rangle$ results,
after some straightforward operations,
\begin{equation}
\langle
n,\tau|\hat{T}^2|n,\tau\rangle\equiv\int_{-\infty}^{\infty}\,dt\,
t^2 \overline{\psi_{n \tau}^{\alpha}(t)}\psi_{n
\tau}^{\alpha}(t)=\tau^2+\int_{-\infty}^{\infty}\,dt\, t^2
\overline{\psi_{n 0}^{\alpha}(t)}\psi_{n 0}^{\alpha}(t),\label{ag}
\end{equation}
where we use $\langle n,0|\hat{T}|n,0\rangle=0$.
\begin{figure}[h]
\begin{center}
\includegraphics{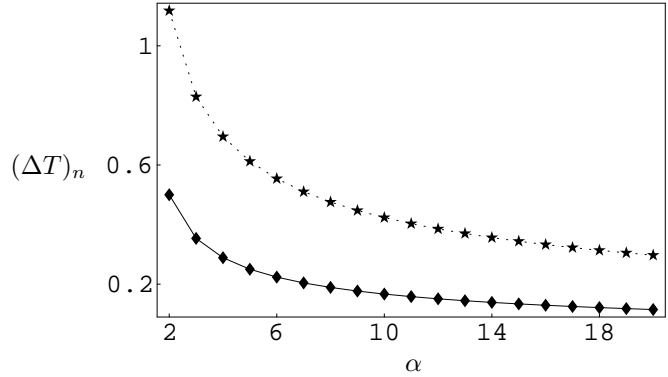}
\caption{$(\Delta T)_n$ (in units of $\omega_0^{-1}$), as a function
of $\alpha$, for $n=0$ (continuous) and $n=3$ (dotted).}
\end{center}
\end{figure}

The last integral in (\ref{ag}) is given in energy representation by
\begin{eqnarray}
\int_{-\infty}^{\infty}\,dt\, t^2 \overline{\psi_{n
0}^{\alpha}(t)}\psi_{n 0}^{\alpha}(t)=-\int_0^{\infty}d\omega
\varphi_n^{\alpha}(\omega)\frac{d^2}{d\omega^2}\varphi_n^{\alpha}
(\omega)=\int_0^{\infty}d\omega[\frac{d}{d\omega}
\varphi_n^{\alpha}(\omega)]^2.\label{ah}
\end{eqnarray}
Making use of (\ref{ae}) we get
\begin{eqnarray}
&&\langle n,\tau|\hat{T}^2|n,\tau\rangle
=\tau^2+\frac{1}{4}+\sum_{l=0}^n(N_{nl}^{\alpha})^2+
\left(\frac{c_n^{\alpha}}{c_{n-1}^{\alpha}}\right)^2\Theta(n,0)\nonumber\\&-&
\sum_{l=0}^nN_{nl}^{\alpha}\int_0^{\infty}d\omega
\varphi_l^{\alpha-2}(\omega) \varphi_n^{\alpha}(\omega)\nonumber\\
&-&2\frac{c_n^{\alpha}}{c_{n-1}^{\alpha}}\Theta(n,0)\sum_{l=0}^n
N_{nl}^{\alpha}\int_0^{\infty}d\omega \varphi_l^{\alpha-2}(\omega)
\varphi_{n-1}^{\alpha}(\omega)\label{ai}
\end{eqnarray}
These expressions can be readily computed. The results confirm the
expectations: The value of $\hat{T}^2$ is near $\tau^2$, the closer
to it the larger $\alpha$. We plot this behaviour in Fig. 3. Notice
that (\ref{ah}) is nothing else but the variance $(\Delta T)_n^2$
for the mode $n$, something explicit after (\ref{ag}).

It is possible to build minimal uncertainty packets by combining
different modes with appropriate coefficients. Instead,  we will
show how close the packets are to the lowest uncertainty. We do
this in Fig. 4. Perhaps, the most prominent feature of these
uncertainty relations is that the lowest uncertainty reached for
each mode is bounded by $n+1/2$, an asymptotic value corresponding
to $\alpha \rightarrow \infty$. On the other hand, the uncertainty
remains bounded for the cases of physical interest.
\begin{figure}[h]
\begin{center}
\includegraphics{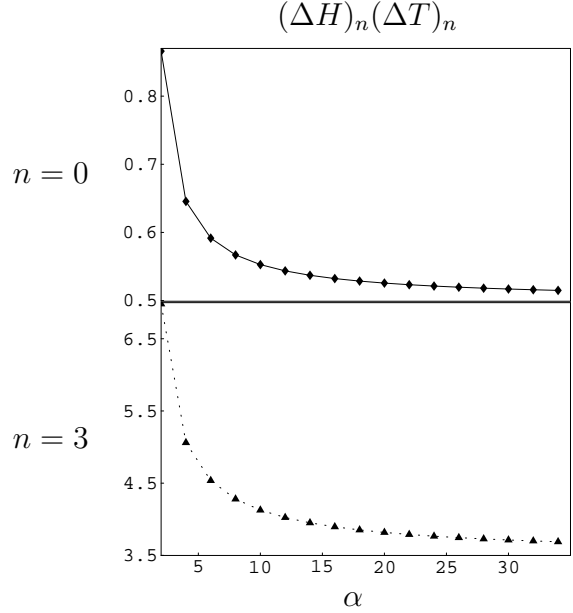}
\end{center}
\caption{$(\Delta H)_n(\Delta T)_n$ as a function of $\alpha$, for
$n=0,3$.}
\end{figure}

Finally, we look for the time of arrival at $x$ of some state
$|\Psi\,\rangle$. We describe the state by a ket to avoid unwieldy
notation, but the discussion could apply equally to density
matrices. Working in parallel to what done to get (\ref{d}) and
(\ref{e}), we could give the modes for mean time of arrival $\tau$
at $x$ with direction $s=\pm 1$ (for right and left movers
respectively). They are given by:
\begin{equation}
\langle n \, \tau \, x\,s\,|\Psi\,\rangle\,= \,\int_0^{\infty}
d\omega\,\langle x \,|\, p(\omega)\,\rangle\, \overline{\varphi_{n
\tau}^{\alpha}(\omega)}\,\langle \omega\,|\Psi\,\rangle \label{aj}
\end{equation}
the notation $p(\omega)$ stands for the dispersion relation
governing the system.
\begin{figure}[h]
\begin{center}
\includegraphics{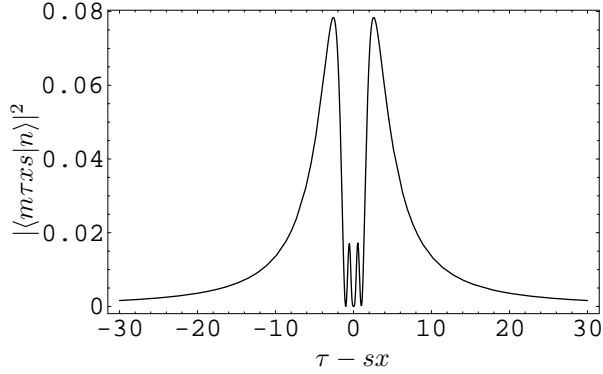}
\end{center}
\caption{$|\langle m \tau x s|n\rangle|^2$ for $m=1$, $n=3$ and
$\alpha=2$.}
\end{figure}
For a massive non relativistic particle $\omega=p^2/2m$, etc. We
choose the case of a massless particle where $p(\omega)=
\,s\,\omega$. This not only avoids square roots but focus on the
very interesting case of photons. In fact for a photon~\cite{iwo1} we could
use $\langle
x\,|\Psi\,\rangle\,=\,\mathbf{F}(x)\,=\,\mathbf{E}(x)\,+\,i\,
\mathbf{B}(x)$, whose Schr\"odinger equation reads
\begin{equation}
i\frac{\partial\mathbf{F}(x,t) }{\partial t}\,=\,-i\,
\left(\mathbf{\sigma}\wedge \mathbf{\nabla}\right)\,
\mathbf{F}(x,t) \label{ak}
\end{equation}
where $\mathbf{\sigma}$ is the photon spin. Admitting the quantum
leap involved in  reducing this to one space dimension, it would
translate into $i\frac{\partial F}{\partial t}\,=\,-i\,
\frac{\partial F}{\partial x}$. This is what we are using for the
dispersion relation. By decomposing the state in terms of the modes
$\langle \omega\,|\Psi\,\rangle\,=\, \sum_m \phi_m^{\alpha}
\varphi_m^{\alpha}(\omega)$, one gets after some computation that
\begin{eqnarray}
\langle n \, \tau \, x\,s\,|\Psi\,\rangle& = & \sum_m
\,\psi_{nm}^{\alpha}
\,\frac{\left(i\omega_0(\tau-sx)\right)^{m+n}}
{\left(1+i\omega_0(\tau-sx)\right)^{m+n+1}}\label{al} \\
&\times&
_2F_1\left(-m,-n;-m-n-\alpha;\frac{1+\left(\omega_0(\tau-sx)\right)^2}
{\left(\omega_0(\tau-sx)\right)^2}\right)\nonumber
\end{eqnarray}
where $\psi_{nm}^{\alpha}=\frac{\omega_0}{\sqrt{2
\pi}}\,c_n^{\alpha}\,c_m^{\alpha}\,\phi_{m}^{\alpha}\,
\frac{\Gamma(m+n+\alpha+1)}{m!n!}$. In Figs. 5 and 6 we plot these
quantities for two cases of interest. Notice the distribution of of
these values around $\tau=sx$. This was expected. Notice also the
zero at this value. This is just the modes orthogonality.
\begin{figure}[h]
\begin{center}
\includegraphics{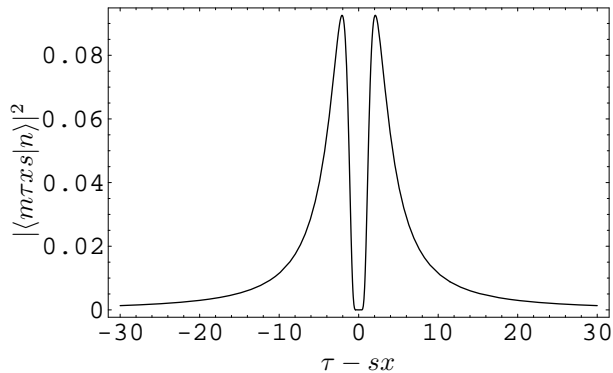}
\end{center}
\caption{$|\langle m \tau x s|n\rangle|^2$ for $m=1$, $n=3$ and
$\alpha=20$.}
\end{figure}

\section*{Conclusions}
We have constructed a representation for time and energy in which
both operate in a selfadjoint manner. The representation has the
nice property of being suitable for probabilistic interpretation. It
is given in terms of a set of modes build in terms of Laguerre
polynomials and their weight functions. This allows to surmount
quite easily the difficulties associated to the different support of
the energy and time spectra. These difficulties not only translate
into the Pauli theorem, but also prevent some asymptotic behaviour
of the modes, precisely the more interesting from the physical point
of view~\cite{iwo2}. According to the Paley Wiener theorem
XII~\cite{paley}, it is not possible to get exponential asymptotic
behaviour for the functions of time that are Fourier transforms of
bounded functions of energy. Here, we dealt with the problem by
using our modes. They are not exponential in time; this is out of
reach. However, they have a controlled variance and uncertainty
relations, in a form that makes them suitable for physics.

\section*{ACKNOWLEDGMENTS}
This work was partially supported by the
Ministerio de Educaci\'on y Ciencia of Spain under project BMF 2002-00834.
The work of L. Lamata was supported by the FPU grant AP2003-0014.

\textit{Note added in proof:} After the submission of this paper the
authors became aware of the interesting article Jos\'e M. Isidro,
Phys. Lett. A 334, 370 (2005), in which a similar problem is studied
from a different approach.

\end{document}